\documentclass[conference]{IEEEtran}

\usepackage{hyperref}
\usepackage{url}
\usepackage{color}
\usepackage{xspace}
\usepackage{graphicx}
\usepackage{amsmath,amsfonts,amssymb}
\usepackage{booktabs}
\usepackage{subcaption}
\usepackage{makecell}
\usepackage{algorithm}
\usepackage{cite}
\usepackage{multirow}
\usepackage{smile}
\usepackage{graphicx}
\usepackage{url}
\usepackage{multirow}
\usepackage{comment}
\usepackage{pifont}
\usepackage{mathtools}
\usepackage{xcolor}
\usepackage{enumitem}
\usepackage{marvosym}
\usepackage{textcomp}
\usepackage{array}
\usepackage{booktabs}
\usepackage{multicol}
\usepackage{balance}
\usepackage{amsmath}
\usepackage{xcolor}
\usepackage{bm}

\usepackage{algorithmic}

\newcommand{\ie}{\textit{i.e.}}

\def\BibTeX{{\rm B\kern-.05em{\sc i\kern-.025em b}\kern-.08em
    T\kern-.1667em\lower.7ex\hbox{E}\kern-.125emX}}
    
\newcommand{\ours}{\sf {TBDS}\xspace}

\begin{document}

\title{Learning Task-Aware Effective Brain Connectivity for fMRI Analysis with Graph Neural Networks}

\author{Yue Yu$^{\dagger}$, Xuan Kan$^\diamond$, Hejie Cui$^\diamond$, Ran Xu$^\diamond$, Yujia Zheng$^\#$, Xiangchen Song$^\ddagger$, Yanqiao Zhu$^\S$, \\
Kun Zhang$^{\#,\ddagger}$, Razieh Nabi$^*$, Ying Guo$^*$, Chao Zhang$^\dagger$, Carl Yang$^\diamond$  \\ 
$^\dagger$College of Computing, Georgia Institute of Technology \\
$^\diamond$Department of Computer Science, $^*$Biostatistics and Bioinformatics, Emory University \\
$^\#$Department of Philosophy, $^\ddagger$Machine Learning, Carnegie Mellon University \\
$^\S$Department of Computer Science, University of California, Los Angeles \\

\texttt{\{yueyu,chaozhang\}@gatech.edu}, \\ \texttt{\{yujiazh,xiangchensong,kunz1\}@cmu.edu}, \\ 
\texttt{yzhu@cs.ucla.edu}, \\ \texttt{\{xuan.kan,hejie.cui,ran.xu,razieh.nabi,yguo2,j.carlyang\}@emory.edu}\\
}

\maketitle

\begin{abstract}
Functional magnetic resonance imaging (fMRI) has become one of the most common imaging modalities for brain function analysis. 
Recently, graph neural networks (GNN) have been adopted for fMRI analysis with superior performance.  
Unfortunately, traditional functional brain networks are mainly constructed based on similarities among region of interests (ROI), which are noisy and  agnostic to the downstream prediction tasks and can lead to inferior results for GNN-based models.  
To better adapt GNNs for fMRI analysis, we propose {\ours}, an end-to-end framework based on \underline{T}ask-aware \underline{B}rain connectivity \underline{D}AG (short for Directed Acyclic Graph) \underline{S}tructure generation for fMRI analysis.
The key component of {\ours} is the brain network generator which adopts a DAG learning approach to transform the raw time-series into task-aware brain connectivities. 
Besides, we design an additional contrastive regularization to inject task-specific knowledge during the brain network generation process. 
Comprehensive experiments on two fMRI datasets, namely Adolescent Brain Cognitive Development (ABCD) and Philadelphia Neuroimaging Cohort (PNC) datasets demonstrate the efficacy of {\ours}. 
In addition, the generated brain networks also highlight the prediction-related brain regions and thus provide unique interpretations of the prediction results. 
Our implementation will be published upon acceptance\footnote{Will be uploaded to https://github.com/yueyu1030/TBDS in the future.}. 
\end{abstract}


\section{Introduction}

{Human brains play a vital role in orchestrating complex neurological systems. 
Understanding the mechanisms of human brains has always been a core interest in the field of neuroscience and valuable to extensive downstream biomedical applications
such as mental disorder therapy~\cite{zhu2021joint}, neural system simulation~\cite{giralt2000simulation}, and cybernetic brain development~\cite{pickering2010cybernetic}. 
Towards this goal, functional magnetic resonance imaging (fMRI) has been acknowledged as a valuable resource of information for brain investigation, which can reflect local changes in cerebral blood oxygenation evoked by sensory, motor, or cognitive tasks~\cite{deyoe1994functional}.  
There has been a significant increase of interest in utilizing fMRI for brain connectome analysis, which focuses on comprehending the brain organizations and their changes, identifying disease-specific biomarkers, as well as supporting clinical decisions such as biological sex prediction~\cite{genderfunction}. }

{To leverage fMRI signals for neurological analysis, traditional biomedical research usually follows a two-stage approach~\cite{smith2012future}. In the first step, \emph{functional brain networks} are generated from blood-oxygen-level-dependent (BOLD) time-series to model the interactions among regions of interests (ROIs). Then, the target classifier is stacked on top of the generated brain networks for downstream clinical predictions~\cite{fcnet,BrainNetCNN,li2020braingnn, kan2022transformer}. Recently, end-to-end neural frameworks have been studied to generate \emph{learnable brain networks} based on embedding similarity and make the prediction simultaneously~\cite{huanghen2019, kan2022fbnetgen}. 
Thus the learned brain networks are more task-oriented under the supervision of task-specific objectives.  
However, two major shortcomings exhibit in both the traditional functional brain networks and the learnable ones. 
Firstly, these brain network generation methods focus on capturing the statistical associations between ROIs. 
Since correlation does not imply causation, they provide insufficient understandings of the complicated brain organization.
Secondly, the connectivity in existing generated brain networks depends on the pairwise similarity between the time-series or embeddings of brain regions, which means that the constructed brain networks are fully or densely connected. 
The noisy signals contained in those dense networks hinder the identification of biological insights on the structure of brain networks and increase the time complexity of the downstream analysis. }

Researchers have proposed a particular type of brain network, effective brain networks~\cite{deshpande2012investigating}, which can overcome these two flaws. This type of brain network aims to infer causal relationships among brain regions and produce sparse connections \cite{stephan2010analyzing}. To construct effective brain networks from BOLD signals, there are several mathematical algorithms available, including Granger causality~\cite{barnett2014mvgc}, dynamic causal modeling~\cite{honey2007network, friston2003dynamic},  autoencoders~\cite{kascenas2022denoising}, and Bayesian search methods~\cite{runge2018causal}. 
However, there are several major drawbacks in directly adopting these techniques for brain connectivity generation tasks:
(1) \emph{Unrealistic assumptions}: these methods often model the brain connectivity with overly simplistic assumptions, such as the absence of unmeasured confounding and lack of temporal dependencies. In reality, such assumptions are hard to satisfy. 
(2) \emph{Limited scalability}: 
existing works based on constraint- or score-based methods for brain connectivity generation~\cite{saetia2021constructing,dubois2020causal} are usually evaluated on a selected ROI subset (less than 50 regions) for their difficulty on scalability. 
But in real application scenarios, there exist hundreds of ROIs, and directly adopting these methods could take several hours, or even several days for each instance. 
(3) \emph{Difficulty of injecting task-specific information}: the above brain network generation methods are not customized for downstream clinical applications~\cite{kascenas2022denoising}. As a result, the mismatch between the network generation and downstream application would hurt the final performance and interpretation.
Till now, generating effective and interpretable brain networks under the supervision of downstream tasks remains a challenging problem.





Fortunately, there is a recent trend in the machine learning community to view structure learning as a directed acyclic graphs (DAG) structure learning problem, which can be further converted to a continuous optimization constrained by additional structural regularizations to ensure acyclicity~\cite{notears, daggnn, golem}. 
Then, this optimization can be solved with some gradient-based approaches, which are efficient, flexible, and can be integrated with other deep learning models. 

Motivated by these studies, we propose {\ours}, a task-aware brain network  generation  approach via 
modeling the connections among different ROIs as DAGs 
to identify effective brain connectivities and predict the target in an end-to-end fashion.   
To tackle the inscalability issue, we leverage the recently proposed approach~\cite{notears,golem} and reformulate the DAG structure learning task as a gradient-based optimization problem,  
which could benefit from GPU acceleration and scales gracefully to hundreds of brain regions.
In addition, to customize the generation process with downstream task knowledge, we design a contrastive loss~\cite{chen2020simple,yu2020fine} to push the brain networks with the same label close and pull the brain networks with different labels apart~\cite{yang2019conditional}. 
Such a regularization enforces the brain networks from different classes to be more distinguishable, so that the downstream GNN classifier can learn to make better decisions.  
In this manner, we can optimize the brain networks towards the downstream tasks, and provide task-specific interpretations to support clinical predictions. 

We evaluate {\ours} on two real-world fMRI benchmarks datasets~\cite{pnc,ABCD} for the important and accessible task of biological sex prediction. 
The results  illustrate that {\ours} achieves competitive performance when compared with advanced baselines. 
Besides, {\ours} is able to characterize the most important brain regions for the target tasks, justifying its efficacy  on providing clinically useful interpretations.

\section{Related Work}
\label{sec:related}
Two lines of the works are relevant to our study.

\subsection{fMRI-based Brain Network Analysis}

Functional Magnetic Resonance Imaging (fMRI) encodes the blood oxygen level dependent (BOLD) signals, and has been widely used to discover the functional connectivity (FC) between different regions in the brain. 
To analyze the functional connectivity with fMRI, traditional methods include 
nonparametric permutation tests~\cite{nichols2002nonparametric,kim2015testing}, 
graph theory based methods~\cite{rubinov2010complex,fornito2013graph} and probabilistic graph modeling approaches \cite{lukemire2021bayesian}. 
However, these methods usually rely on hand-crafted features to infer the functional connectivity, and the generated brain networks are task-agnostic.
With the rapid development of graph neural networks (GNNs), much attention has been paid to \emph{GNN models for fMRI-based brain network analysis}~\cite{li2020braingnn,cui2022interpretable}.  
One major advantage of GNN-based models is that they are capable of aggregating node features based on graph structures and can be trained in an end-to-end fashion to solve downstream tasks.

However, GNNs usually require explicitly given graph structures and node attributes, which are often absent for fMRI analysis.
To tackle this issue, some works directly use the brain networks constructed manually based on statistical correlations \cite{modellingfmri}, but one drawback is that they cannot well handle the negative weights. 
Moreover, the statistical correlations are not specific to the downstream prediction tasks. 
Recently, there are some works that attempt to generate the brain networks and predict for the downstream tasks jointly~\cite{a14030075,kreuzer2021rethinking,kan2022fbnetgen}. 
But they are mainly based on structure similarity from GNNs~\cite{kan2022fbnetgen} or attention weights~\cite{a14030075,kreuzer2021rethinking}, which still cannot well characterize the complex relationships among different regions.  




\subsection{DAG Structure Learning}
Learning graphical structures based on Directed Acyclic Graphs (DAGs) is a fundamental problem in machine learning with a broad range of applications such as healthcare~\cite{lucas2004bayesian,xu2022precise}, social networks~\cite{nguyen2012influence} and biomedical informatics~\cite{siddiqi2022causal}.
Traditional structure learning methods include constraint- and score-based methods, which test the validity of the generated graph with pre-defined scoring functions~\cite{chickering2002optimal,penny2012comparing} or constraints~\cite{spirtes2000causation,maxwell1997efficient}. However, one major drawback of such methods is the \emph{computational inefficiency}, as the search problem is
NP-hard~\cite{Woeginger2003} and the number of possible DAGs often increases super-exponentially with the number of nodes in the graph.

To reduce the computational overhead, \cite{notears} recasts the DAG search problem as a purely continuous optimization problem, where an additional penalty term is introduced to enforce the acyclicity of the generated graph. After that, many other works have extended this line of research~\cite{golem,pmlr-v108-pamfil20a} to improve it both theoretically and empirically.   
Despite its effectiveness, DAG structure learning approaches are largely unexplored by current research for fMRI data. 
As recent works demonstrate the existence of causal links among different ROIs~\cite{friston2009causal,hill2017causal,siddiqi2022causal}, it is possible to leverage structure learning approaches to reveal the relations among different ROIs. 



\begin{figure}[t]
    \centering
    \includegraphics[width=0.92\linewidth]{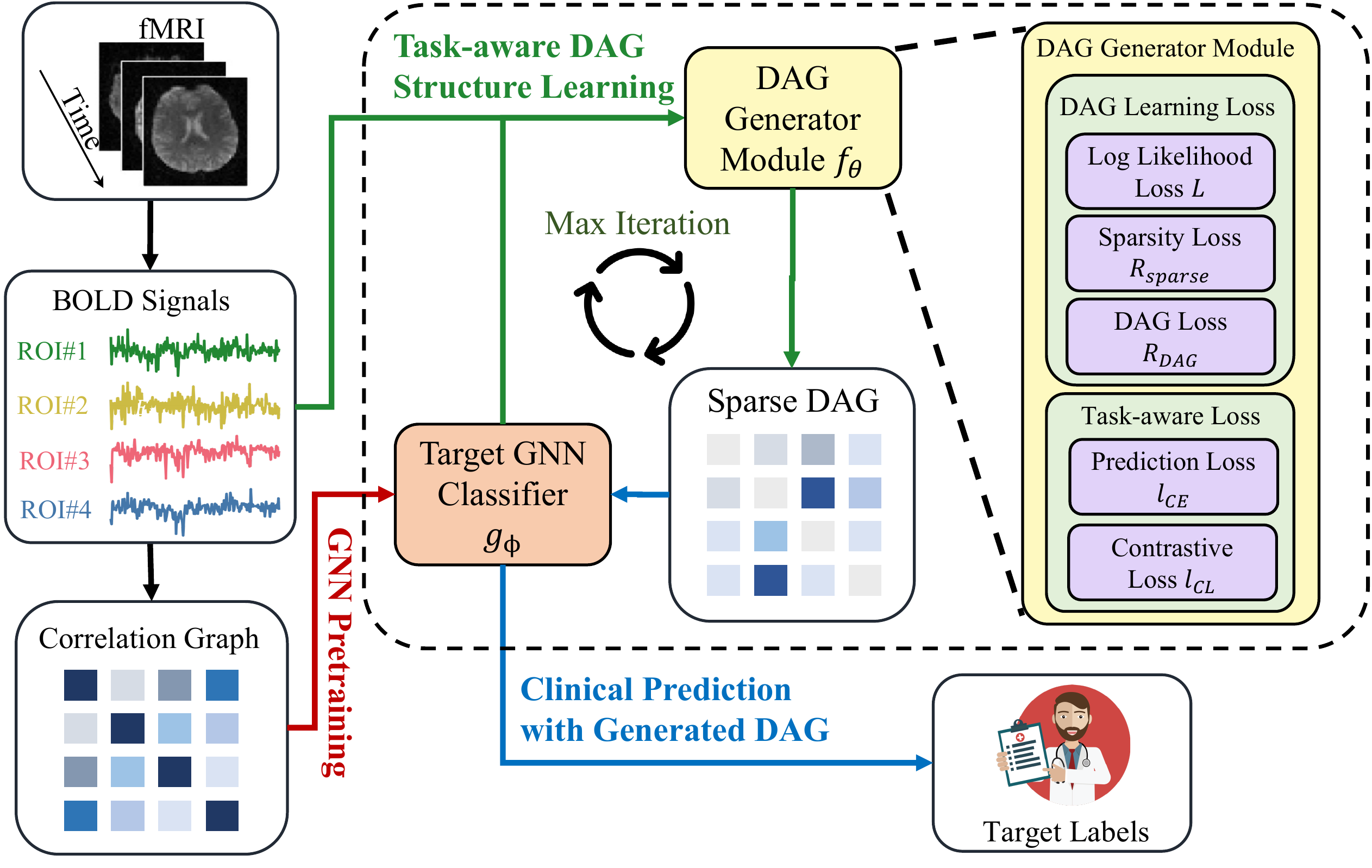}
    \caption{The framework of {\ours}. Different colors indicate different training stages.}
    \label{fig:model}
\end{figure}

\section{Method}
\label{sec:method}
In this section, we first briefly summarize the task studied in this paper, then introduce the pipeline of using graph neural networks for prediction tasks on brain networks. Finally, we introduce how to leverage task-aware DAG learning techniques to generate effective brain connectivities. 
\subsection{Task Definition.} 
In this study, the input $\bm X\in \mathbb{R}^{n \times v \times t}$ is the BOLD time-series for regions of interest (ROIs), where $n$ is the sample size, $v$ is the number of ROIs and $t$ is the length of time-series. 
The target output for each time-series consists of two terms: 
(1) a functional brain network $\bm A\in\mathbb{R}^{v\times v}$ (\ie, brain connectivity matrix) for each sample $\bm x\in\mathbb{R}^{v\times t}$, which serves as an intermediate result to the end-to-end framework.
(2) the prediction $\bm Y \in \mathbb{R}^{n \times |\mathcal{C}|}$, where $|\mathcal{C}|$ is the number of classes. 

\subsection{GNN for Brain Networks} 
Graph neural networks (GNNs) have been widely used in fMRI-based brain network analysis~\cite{li2020braingnn,cui2022braingb, yang2022data} 
due to their strong ability in learning node representations via propagating neighbor node features.  
However, one of the major challenges for adopting GNNs for brain networks is that they cannot  handle edges with \emph{negative weights} well. 
In this study, we leverage the \emph{modified} $k$-layer graph convolutional network architecture~\cite{kipf2017semisupervised}  to accommodate negative edges in $\bA$ for clinical prediction tasks\footnote{More complicated GNN structures can also be integrated with {\ours}, which is of separate interests.}.
Let $\bm{\Theta}^{(k)} = (\bm\theta_1^{(k)}, \bm\theta_2^{(k)}, \ldots, \bm\theta_{v}^{(k)})$ be the matrix of all node embedding vectors at step $k$, the update rule for node embedding is 
\begin{equation}
    \bm{\Theta}^{(k)} = \operatorname{ReLU}\left(\hat{\bm{A}}_{+} \bm{\Theta}^{(k-1)} \bW_{+}^{(k)} + \hat{\bm{A}}_{-} \bm{\Theta}^{(k-1)} \bW_{-}^{(k)}\right),
\label{eq:gcn}
\end{equation}
where $\hat{\bA}_{+}=\bD_{+}^{-\frac12}\left(I+\operatorname{ReLU}(\bA)\right)\bD_{+}^{-\frac12}$ and $\hat{\bA}_{-}=\bD_{-}^{-\frac12}\left((I+\operatorname{ReLU}(-\bA))\right)\bD_{-}^{-\frac12}$ stands for the normalized adjacency matrices for positive edges and negative edges.  
$\bD_{+} = \operatorname{diag}\left(\sum_j\operatorname{ReLU}(\bA)_{:,j}\right)$, $\bD_{-} = \operatorname{diag}\left(\sum_j\operatorname{ReLU}(-\bA)_{:,j}\right)$ is two diagonal matrices representing node degrees. 
$\bW^{(k)}$ stands for learnable parameters in convolutional layers and $\bm{\Theta}^{(0)}=\bm F$ is the \emph{connection vector} that stands for Pearson correlation scores between its time-series with all the nodes contained in the brain network, which is suggested by~\cite{cui2022braingb}.  
After obtaining the node embeddings from the last GNN layer $\bm{\Theta}^{(k)}$, we concatenate all the node embedding and use an additional classification head for the target prediction as 
\begin{align}
    \hat{\by} &=  \sigma\left(\operatorname{MLP}\left(\operatorname{BatchNorm1D}(\|_{j = 1}^{v} \bm{\theta}^{(k)}_j)\right)\right); \label{eq:gcn_pred} \\
    \cL_{\text{tgt}} &= \ell_{\text{CE}}(\hat{\by}, y),
    \label{eq:ce_loss}
\end{align}
where $\hat{\by}$ is the probability simplex of the prediction and $\sigma(\cdot)$ is the softmax function.

\subsection{Task-aware DAG Structure Learning from BOLD Signals} 
\noindent $\diamond$ \textbf{Formulation.} For the input BOLD time-series $\bm x$, generating DAGs to capture the effective connectivities is not easy due to the complex spatio-temporal relationships among the ROIs: the current signal of a brain region is related to previous signals from both itself and other ROIs~\cite{mitra2014lag}. 
In this work, we hypothesize that different ROIs influence one another in  both contemporaneous and time-lagged manner. 
Then, we harvest the standard SVAR model~\cite{swanson1997impulse} to  consider both  spatial and temporal relationships during the DAG generation process. In the $t$-th step, the SVAR model can be expressed as 
\begin{equation}
\bm{x}_{t} =\bm{A}^{\top}\bm{x}_{t} +\bm{A}_{1}^{\top}\bm{x}_{t-1}  +\ldots+\bm{A}_{p}^{\top}\bm{x}_{t-p} +\bm{\epsilon}_t. 
\label{eq:svar}
\end{equation}
To interpret Eq.~\ref{eq:svar}, we note that $\bm{x}_t$ is an $v$-dimensional vector that represents the signal of all ROIs at the $t$-th step, $\bm{\epsilon}_t$ is a noise vector with independent elements~\cite{notears}, and $p$ is the autoregressive order. 
Besides, $\bm{A}_i$ $(i=1,2,\ldots,p)$ represent weighted adjacency matrices with nonzero entries corresponding to the time-lagged relations, and $\bm{A}$ is the directed acyclic graph (DAG) to model the effective relations among ROIs. 
To effectively model the BOLD signals among different time steps, we further assume the DAG structure is time-invariant and transform Eq.~\ref{eq:svar} into the matrix form as 
\begin{equation}
\bm{X} =\bm{A}^{\top}\bm{X} +\bm{A}_{1}^{\top}\bm{X}_1 +\ldots+\bm{A}_{p}^{\top}\bm{X}_p +\bm{E},  
\label{eq:svar_mt}
\end{equation}
where $\bm{X}$ is a $v\times(T+1-p)$ matrix with each column stands for one timestep and  $\bm{X}_1, \ldots, \bm{X}_p$ are time-lagged versions of $\bX$. 
Denote $\bZ = \left[\bm{X}_{1}|\cdots| \bm{X}_{p}\right]$ and  $\bm{B}=\left[
\mathbf{A}_{1}^{\top} |\cdots| \mathbf{A}_{p}^{\top}\right]^{\top}$, we obtain the compact form of Eq.~\ref{eq:svar_mt} as
\begin{equation}
\bm{X} =\bm{A}^{\top}\bm{X} +\bB^{\top}\bZ +\bm{E}.  
\label{eq:svar_matrix}
\end{equation}

\noindent $\diamond$ \textbf{Optimization Objective for Vanilla DAG  Learning.}  To learn the DAG structure for brain networks with $\bX, \bZ$, we aim to estimate the matrix $\bA$ and $\bB$ to satisfy Eq.~\ref{eq:svar_mt} as well as the directed acyclic constraint (for $\bA$ only). 
This indeed creates a constraint optimization problem, written as follows 

\begin{small}
\begin{equation}
\min_{\bm{A}, \bm{B}} \ \  \ell(\bm{A}, \bm{B})=\frac{1}{2}\|\bm{X}-\bm{A}^{\top}\bm{X} -\bB^{\top}\bZ\|_{{F}}^2  \text { \ s.t. \ } \bm{A} \text { is acyclic.}
\label{eq:opt_re}
\end{equation}
\end{small}
Here 
$
\ell(\bm{A}, \bm{B})
$
is the objective for DAG generation.

Directly optimizing Eq.~\ref{eq:opt_re} is difficult due to the hard  acyclicity constraint~\cite{notears}.
To circumvent this issue, we follow the recent work~\cite{golem} and reformulate the above task to an \emph{unconstrained optimization problem}
via applying additional \emph{soft sparsity} and \emph{acyclic} regularization for learning a DAG equivalent to the ground truth DAG.
The overall objective function is defined as 

\begin{small}
\begin{align}
\min_{\bA, \bB} \ \mathcal{S}(\bA, \bB  ; \bm{X})&=\mathcal{L}(\bA, \bB ; \bm{X}) \label{eq:log_likelihood}\\
&+\lambda_1 \left(\cR_{\text{sparse}}(\bA)+\cR_{\text{sparse}}(\bB)\right)  \\
&+\lambda_2 \cR_{\text{DAG}}(\bA) 
\label{eq:unconstrain}
\end{align}
\end{small}
where
\begin{small}
\begin{align}
\mathcal{L}(\bA, \bB ; \bm{X}) &=-\frac{1}{2} \sum_{i=1}^{v} \log \left(\sum_{j=1}^{T+1-p}\left(\bX_{i,j}-\bA_{i}^{\top} \bX_{:, j}-\bB_{i}^{\top} \bZ_{:, j}\right)^{2}\right) \\
&+\log |\operatorname{det}(I-\bA)| +\text{const.}
\end{align}
\end{small}
$\mathcal{L}(\bA, \bB ; \bm{X})$ is the learning objective of the maximum log-likelihood estimator of $\bX, \bZ$ following multivariate Gaussian distribution.
In addition, we use $l_1$ penalty to approximate the sparse constraint and use the DAG constraint proposed in~\cite{notears} as the realization of $\cR_{\text{DAG}}$ which can be  written as 
\begin{equation}
\cR_{\text{sparse}}(\bA) = \|\bA\|_{1},  \quad
\cR_{\text{DAG}}(\bA) = \operatorname{tr}\left(e^{\bA \circ \bA}\right)-v.
\end{equation}
By optimizing $\bA$ and $\bB$ with the above equations, we are able to obtain the brain network $\bA$, which will be also used in the downstream task described as follows.


\noindent $\diamond$ \textbf{Task-aware DAG Structure Learning.}  
\label{ref:graph_gen}
The previous sections have not fully explore the connections between the DAG generation model as $\bA = f_\theta(\bX)$, and the target prediction model as $\hat{\by} = g_{\phi}(\bA, \bF)$. 
In this work, we aim to establish the connection between $f_\theta(\bX)$ and  $g_{\phi}(\bA, \bF)$ by leveraging $g_{\phi}(\bA, \bF)$ to guide the structure learning process.   

To achieve this, we first pretrain the $g_{\phi}$ on the training set by setting $\bA=\bF$. In this way, the brain connectivity is initialized with the Pearson correlation matrix, which reflects statistical correlations among nodes. Such connectivities could serve as a warmup for the training of $g_{\phi}$. We denote the pretrained model as $g_{\phi}^{(0)}$.
 With an estimated target classifier from $g_{\phi}$, {\ours} trains the generator $f_\theta(\bX)$ to generate \emph{more discriminative} DAGs easier for  $g_{\phi}(\bA, \bF)$ to classify.
The learning objective of the task-aware DAG structure learning can be written as 
\begin{equation}
\begin{aligned}
\min_{\bA, \bB} \ \cL_{\text{Gen}} &=  \mathcal{S}(\bA, \bB ; \bm{X})  \\
&+ \mu \left(\ell_{\text{CE}}\left(g_{\phi}(\bA, \bF), y\right) + \ell_{\text{CL}}\left(g_{\phi}(\bA, \bF), y\right) \right), 
\end{aligned}
\label{eq:contrast_gen}
\end{equation}
where $\mu$ is a hyperparameter to control the weight of two regularization terms, and $\ell_{\text{CL}}$ is the contrastive loss. 
Note that in this step, we keep $g_{\phi}$ \emph{fixed} without updating its parameters.  
To calculate  $\ell_{\text{CL}}$, we first calculate the representation for each DAG as $\bv=\text{GNN}(\bA)\in \mathbb{R}^{d}$ (the architecture of GNN is in Eq.~\ref{eq:gcn}.). Then, we use the margin-based contrastive loss as 

\begin{small}
\begin{equation}
\begin{aligned}
&\ell_{\text {pos},i}=\frac{\sum_{p \in \cP_i}\left\|\boldsymbol{v}_{i}-\boldsymbol{v}_{p}\right\|^{2}}{|\cP_i|},  \ell_{\text {neg},i}= \frac{\sum_{n \in \cN_i}\left(\xi-\left\|\boldsymbol{v}_{i}-\boldsymbol{v}_{n}\right\|^{2}\right)_{+}}{|\cN_i|} , \\
&\ell_{\text{CL}}=\sum_{i=1}^{M} \frac{1}{dM}\left(\ell_{\text {pos},i}+\ell_{\text{neg},i}\right).
\end{aligned}
\end{equation}
\end{small} 
Here $M$ is the number of DAGs in a batch, $\cP_i$ and $\cN_i$ is the set of instances from the same or different class from $i$-th instance, respectively. 

We remark that $\ell_{\text{CE}}$ and $\ell_{\text{CL}}$ optimize the DAG generation from different perspectives: $\ell_{\text{CE}}$ directly optimizes over the final prediction, while $\ell_{\text{CL}}$ operates on the representation space to push the embeddings of DAGs from the same class close and pull the embeddings from different classes further apart~\cite{yu2020fine}. 
By injecting these two terms, we are able to generate task-specific DAGs to better support the clinical prediction tasks.


\begin{figure}[t]
	\label{alg:LSB}
	\renewcommand{\algorithmicrequire}{\textbf{Input:}}
	\renewcommand{\algorithmicensure}{\textbf{Output:}}
	\begin{algorithm}[H]
		\caption{The training procedure of {\ours}.}
		\begin{algorithmic}[1]
			\REQUIRE BOLD time series $\bX$, Pearson Correlation Matrix $\bF$          
			\ENSURE Brain Network $\bA$, prediction $\hat{y}$, DAG generation module $f_{\theta}$, target GNN classification module $g^{0}_{\phi}$.   
			\STATE $i$ = 0.
			\STATE // \textit{Training Stage}
			\STATE Pretrain $g_{\phi}$ via setting $\bA=\bF$ (Eq.~\ref{eq:gcn}-~\ref{eq:ce_loss}).
			\WHILE {$i < \text{MaxIter}$}
			\STATE Generate $\bA^{(i)}, \bB^{(i)}$ on the training set using $g^{(i-1)}_{\phi}$ (Eq.~\ref{eq:contrast_gen}).
			\STATE Train $g^{(i)}_{\phi}$ with the generated $\bA^{(i)}$ (Eq.~\ref{eq:gcn}-~\ref{eq:ce_loss}).
			\STATE $i = i + 1$.
			\ENDWHILE
			\STATE // \textit{Inference Stage}
			\STATE  Generate $\bA, \bB$ on the test set with pseudo labels derived from $g^{(i)}_{\phi}$  (Eq.~\ref{eq:contrast_gen_test}).
			\STATE Inference the labels $\hat{y}$ on the test set with $g^{(i)}_{\phi}$ (Eq.~\ref{eq:gcn}-~\ref{eq:gcn_pred}).
		\end{algorithmic}
			\label{alg:main_1}
	\end{algorithm}
\end{figure}

\noindent $\diamond$ \textbf{Inference on Test Data.} 
One challenge for the method mentioned above is that for \emph{test data}, we do not have any ground truth label $y$, thus directly using Eq.~\ref{eq:contrast_gen} is infeasible. 
To tackle this issue, we propose to leverage the \emph{pseudo labels} predicted from $\tilde{y}=\arg \max (\hat{\by})$ as a substitute of the ground truth labels.  Then, to suppress the label noise in pseudo labels, we add an additional reweighting term via leveraging the predictive uncertainty $\omega_{\bA}=1-\operatorname{Entropy}(\hat{\by})=1-\sum_i \hat{\by}_i \log{\hat{\by}_i}$. 

In this way, if the model is \emph{certain} about its prediction, the entropy would be low, and we assign higher weight for two additional terms to encourage the model to learn from the target tasks. 
In contrast, when the model prediction is uncertain indicating that it is more possible to be errorneous, we reduce the weight to prevent it from overfitting to the noise. The overall learning objective for test data is expressed below.

\begin{small}
\begin{equation}
\begin{aligned}
\min_{\bA, \bB} \ \cL_{\text{Gen,test}} &=  \mathcal{S}(\bA, \bB ; \bm{X})  \\
&+ \mu \cdot \omega_{\bA} \left(\ell_{\text{CE}}\left(g_{\phi}(\bA, \bF), \hat{y}\right) + \ell_{\text{CL}}\left(g_{\phi}(\bA, \bF), \hat{y}\right) \right), 
\end{aligned}
\label{eq:contrast_gen_test}
\end{equation}
\end{small}

\subsection{End-to-end Training} 
The overall procedure of {\ours} is shown in Alg. \ref{alg:main_1}. 
It is worth noting that the two modules $f_{\theta}$ and $g_{\phi}$ are trained in an end-to-end way, as the label $y$ and the task-oriented graphs are leveraged simultaneously.
Moreover, the  brain networks  of the test set are inferred \emph{separately} from the training set. Thus, we eliminate the issue of information leakage, as \emph{no information} from the test set has been used during the tuning process.

\section{Experiments}
\label{sec:exp}
In this section, we conduct extensive experiments to answer the
following three research questions: 
\textbf{RQ1}: How does {\ours} perform as compared with state-of-the-art methods? 
\textbf{RQ2}: How do the key designs in {\ours} affect performance? 
\textbf{RQ3}: Does the generated brain connectivity by {\ours} offers reasonable interpretability for target prediction tasks?


\subsection{Experiment Setup}
\noindent $\diamond$ \textbf{Dataset.} We conduct experiments demonstrating the utility of {\ours} using two real-world fMRI datasets. 

(a) \textit{Philadelphia Neuroimaging Cohort (PNC)} is a dataset curated from a collaborative project from the Brain Behavior Laboratory at the University of Pennsylvania and the Children's Hospital of Philadelphia. It includes a population-based sample of individuals aged 8–21 years \cite{pnc}. After quality control, we utilize rs-fMRI data of 503 subjects, with 289 (57.46\%) of them being females. 
The nodes are grouped into 10 functional modules that correspond to major resting-state networks. 
In the resulting data, each sample contains 264 nodes with time-series data collected through 120 time steps. 

(b) \textit{Adolescent Brain Cognitive Development Study (ABCD)} \cite{ABCD} is one of the largest publicly available fMRI datasets. 
This study is recruiting children aged 9–10 years across 21 sites in the U.S. Each child is followed into early adulthood, with repeated imaging scans as well as extensive psychological and cognitive testing.  
After quality control, the dataset include fMRI data  of 7901 children, and 3961 (50.1\%) among them are female. 
After processing, each sample contains 360 nodes. Only samples with at least 512 time points are selected, and the first 512 time points are included in the dataset.

\noindent $\diamond$ \textbf{Task.} We choose biological sex prediction as the evaluation task since 
sexuality is an essential aspect of adolescent development, biological sex prediction is a critical and meaningful task for ABCD and PNC. Many papers \cite{POTTER2022101057, genderfunction,kan2022fbnetgen} have focused on this task using brain networks.


\noindent $\diamond$ \textbf{Evaluation Metrics.} 
As the label distributions of both PNC and ABCD datasets are balanced, we use both \emph{AUROC} and \emph{accuracy} as the performance metrics. For accuracy, we  use $0.5$ as the threshold after obtaining the predicted result.

\noindent $\diamond$ \textbf{Implementations.}  We implement our model in PyTorch\footnote{\url{https://pytorch.org/}}. 
We use Adam as the optimizer with the learning rate 1e-4. The key hyperparameters in {\ours} include regularization weight $\lambda_1, \lambda_2, \mu$ and the number of steps $p$ in SVAR model.   
Following common practice~\cite{golem}, we set $\lambda_1=1, \lambda_2=10, \mu=10$ without further tuning. Following~\cite{saetia2021constructing}, we set $p=3$ as further increasing $p$ will introduce more learnable parameters  which causes OOM error for large datasets (ABCD). 
MaxIter is set to 3 for PNC and 2 for ABCD as we find increasing the number iteration will not significantly increase the model's performance but extend the running time. 

\subsection{Baselines} We compare {\ours} with the following baselines:
\footnote{To ensure fair comparison, we tune the parameter for {\ours} and  baselines and choose the best one based on the performance of the development set.}

\textbf{(a) Time-series based Models.} These baselines model BOLD time-series data \emph{without} networks. Here we select \emph{bi-GRU} as the baseline to encode BOLD time-series. 

\textbf{(b) Statistical Methods for Brain Networks.}
These baselines use statistical methods to construct brain networks, including \emph{uniform graphs}, which sets all entries in adjacency matrices with one and \emph{Pearson correlation graphs}, which sets the weight in adjacency matrices as Pearson correlation of BOLD signals. To enable fair comparison, 
we use the same node features and GNN predictor as {\ours} for prediction.

\textbf{(c) Deep Learning Models for Brain Networks.} We also compare our method with four popular deep models for brain networks, \emph{BrainnetCNN}~\cite{BrainNetCNN}, \emph{BrainGNN}~\cite{li2020braingnn},  \emph{SAN}~\cite{kreuzer2021rethinking}, and \emph{BrainGB}~\cite{cui2022braingb}. 
Note that these methods develops advanced neural networks with the fixed correlation-based functional brain networks to model the relations among different ROIs.

\textbf{(d) Models with Learnable Brain Network Generation.} We introduce three other baselines based on learnable graph generators, namely \emph{LDS}~\cite{Learninggraph2019}, \emph{FBNetGen}~\cite{kan2022fbnetgen} and \emph{GNN-DAG}~\cite{saetia2021constructing}. LDS is a framework which jointly learns graph structures and model parameters through bilevel optimization.  
FBNetGen is an end-to-end approach for learning task-aware brain networks  based on embedding similarities.  
GNN-DAG leverages causal discovery algorithm~\cite{runge2018causal} for generating brain networks. Note that the original GNN-DAG method is not designed for downstream tasks, and we use the same GNN classifier to adapt GNN-DAG for target tasks. 






\subsection{Main Experiment (RQ1)}
\begin{table*}[tbp]
\centering
\renewcommand\arraystretch{0.95}
\small
\label{tab:performance}
\resizebox{0.72\linewidth}{!}{
\begin{tabular}{cccccccc}
\toprule
\multirow{2.5}{*}{Type} & \multirow{2.5}{*}{Method} &\multicolumn{2}{c}{\bf Dataset: PNC} & \multicolumn{2}{c}{\bf Dataset: ABCD}\\
\cmidrule(lr){3-4} \cmidrule(lr){5-6} 
& & {AUROC} & {Accuracy} & {AUROC} & {Accuracy}  \\
\midrule
\multirow{1}{*}{Time-series} 
& bi-GRU &  65.1 ± 3.5 & 58.1 ± 2.4 & 51.2 ± 1.0 & 49.9 ± 0.8  \\
\midrule
\multirow{2}{*}{\shortstack{Traditional \\ Graph}} 
& GNN-Uniform & 70.6 ± 4.8 & 66.2 ± 3.9 & 88.8 ± 0.7 & 80.5 ± 0.7 \\
& GNN-Pearson & 76.5 ± 2.7 & 69.2 ± 1.8 & 91.0 ± 0.5 & 82.4 ± 0.5 \\
\midrule
\multirow{5}{*}{Deep Model} 
& BrainnetCNN~\cite{BrainNetCNN}  & 78.5 ± 3.2 & 71.9 ± 4.9  & 93.5 ± 0.3 & 85.7 ± 0.8 \\
& BrainGNN~\cite{li2020braingnn} & 77.5 ± 3.2 & 70.6 ± 4.8  & OOM & OOM\\
& SAN~\cite{kreuzer2021rethinking} & 73.0 ± 1.6 & 70.4 ± 2.4 &  90.1 ± 1.2 & 81.0 ± 1.3 \\
& BrainGB~\cite{cui2022braingb} & 76.6 ± 5.0	& 69.8 ± 4.2 & 91.8 ± 0.3 & 83.1 ± 0.9  \\
\midrule
\multirow{3}{*}{\shortstack{Learnable \\ Graph}} 
& LDS \cite{Learninggraph2019} & 78.2 ± 3.8 & 70.8 ± 6.2  & 90.7 ± 0.3 & 82.5 ± 0.9 \\
& FBNetGen~\cite{kan2022fbnetgen} & 80.8 ± 3.3  & 74.8 ± 2.3  & \textbf{94.5 ± 0.7}  & 87.2 ± 1.2 \\ 
& GNN-DAG \cite{saetia2021constructing} & 80.5 ± 2.6  & 72.3 ± 2.0 & $>$ 1 week& $>$ 1 week \\
\midrule
\multirow{1}{*}{Ours} 
& {\ours} & \textbf{83.4 ± 1.9} & \textbf{76.9 ± 1.8} & {94.2 ± 0.2} & \textbf{88.0 ± 0.4}  \\
\bottomrule
\end{tabular}
}
\caption{Performance (in \%) comparison with different types of baselines. Note that \textbf{OOM} means out-of-memory error, and \textbf{$>$1 week} means the algorithm cannot be finished in 1 week. PNC is a \emph{low-resource} dataset including 503 fMRI data,  ABCD is a \emph{high-resource} dataset including 7901 fMRI data. }
\label{tab:main_result}
\end{table*}

\begin{figure}
	\centering
	\subfloat[AUROC]{
		\includegraphics[width=0.42\linewidth]{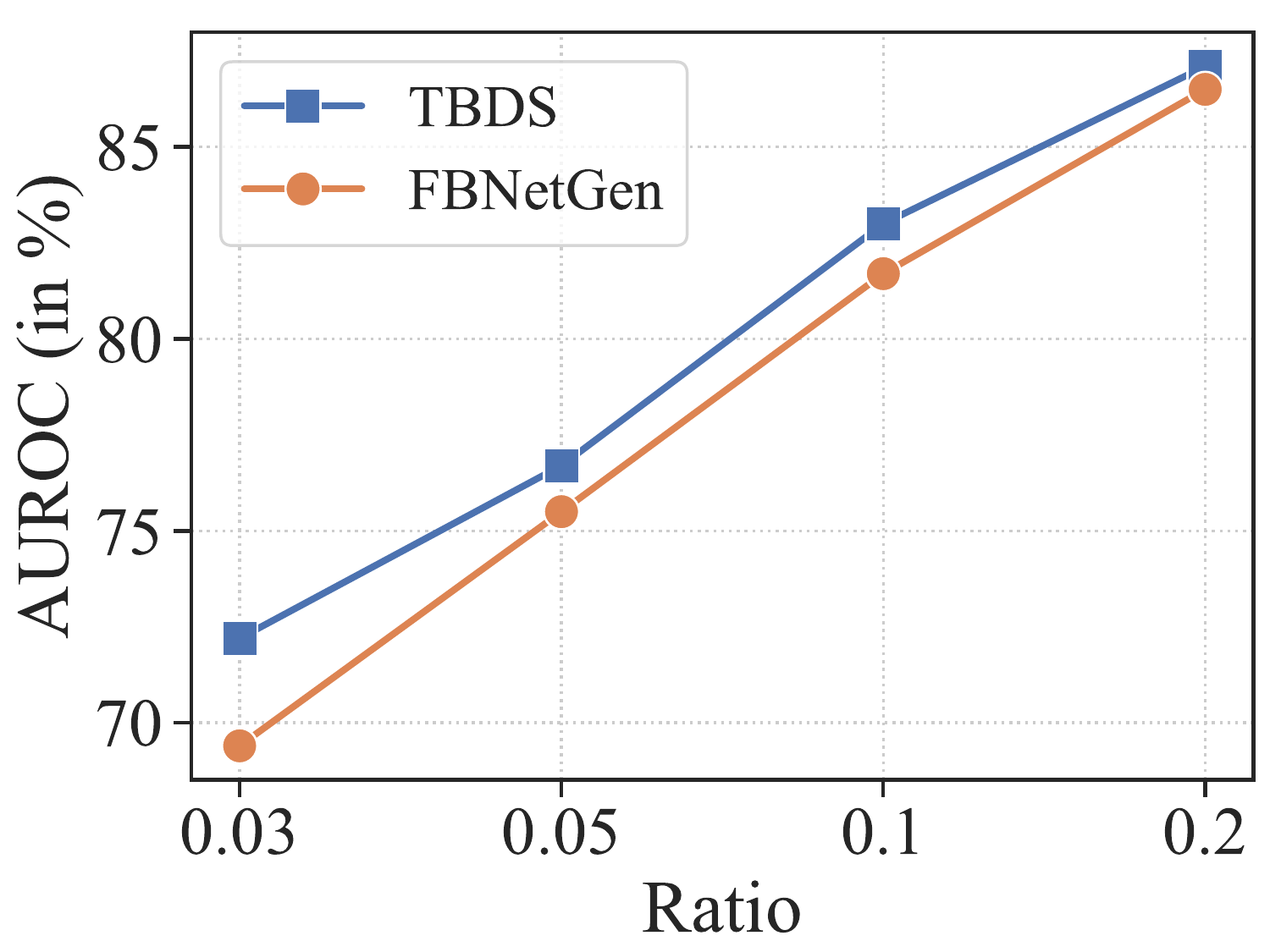}
		\label{fig:auc}
	} 
	\subfloat[Accuracy]{
		\includegraphics[width=0.42\linewidth]{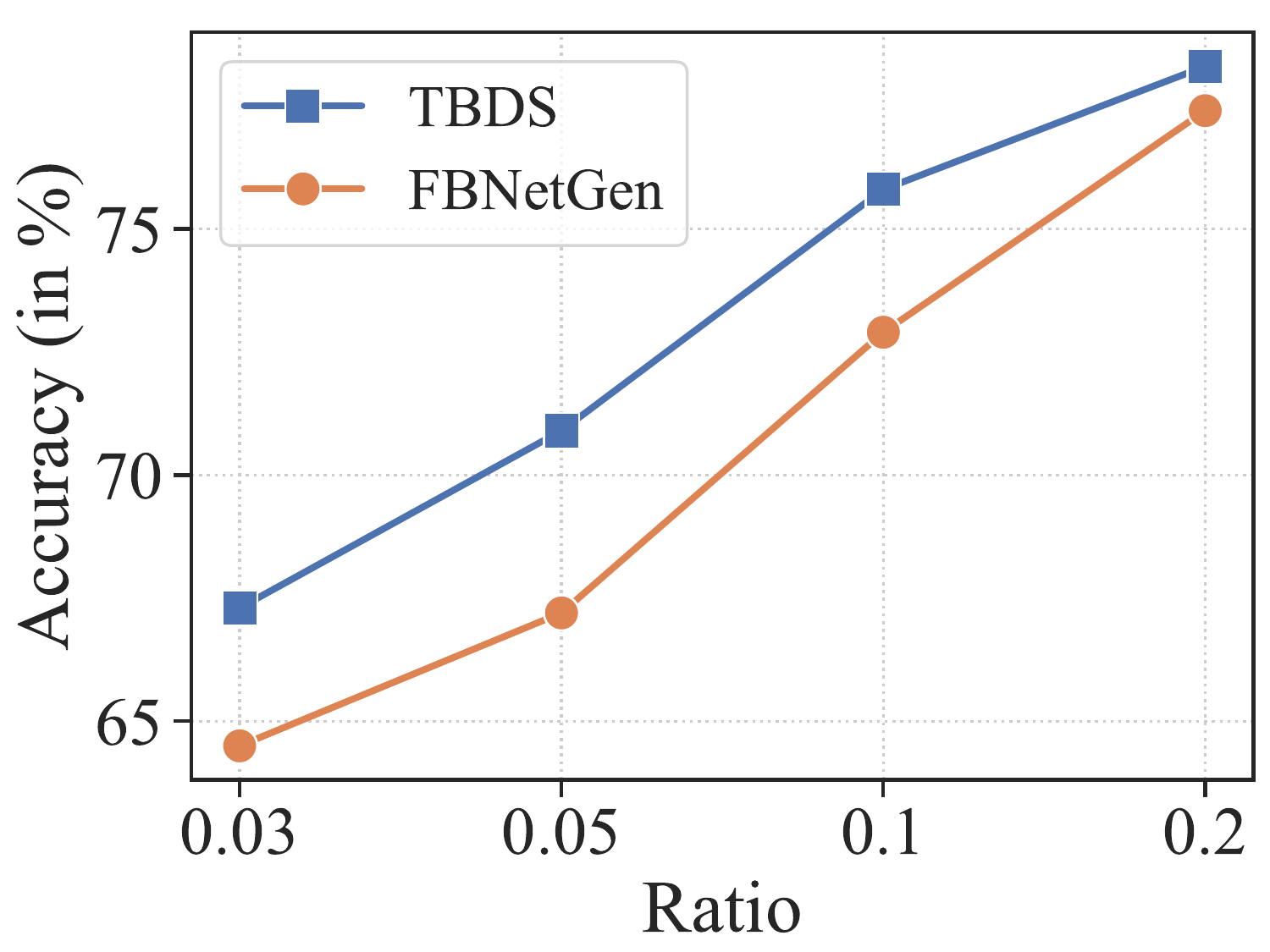}
		\label{fig:acc}
	}
	\vspace{-1ex}
	\caption{Performance of {\ours} and FBNetGen trained with different fractions of labeled data in ABCD training set.}
	\vspace{-1ex}
\label{fig:fraction}
\end{figure}

From the experimental results shown in Table~\ref{tab:main_result}, we have the following findings:
\begin{itemize}[leftmargin=*]
    \item {\ours} outperforms all the baselines for the PNC dataset with 3\% performance gain. On the ABCD dataset, it achieves competitive performance when compared with the strongest baseline (FBNetGen). This is because ABCD is a much larger dataset (16 times larger than PNC) with plenty of labels, and directly using embedding similarity is sufficient for capturing the relations among ROIs. Conversely, when labeled data is limited, FBNetGen cannot perform as well as {\ours}. To further validate this point, we run {\ours} and FBNetGen on a subset of ABCD with different fraction of labeled data (\{3\%, 5\%, 10\%, 20\%\}). As shown in Fig.~\ref{fig:fraction}, {\ours} continuously outperform FBNetGen, which justifies the efficacy of {\ours} under low-data regime. 
    \item Directly using statistical correlations as brain connectivity is insufficient to capture the relationships among ROIs, as they achieve suboptimal results on two datasets. Designing more complicated GNN architectures does not address this challenge --- baselines with carefully designed  GNN models~\cite{BrainNetCNN,cui2022braingb,kreuzer2021rethinking} still underperform {\ours} on two benchmarks. 
    \item Previous methods leveraging DAG generation approaches can outperform statistical-based brain network generation methods. However, the biggest obstacle is the  time complexity --- generating brain networks on larger datasets often takes too much time. Such an inefficiency issue hinders it from being deployed in real applications. In {\ours}, we use continuous optimization techniques with GPU acceleration, which reduce the overall running time significantly (1.5 hours for PNC and 5 hours for ABCD).
\end{itemize}




\begin{table}[!tbp]
\centering
\small
\resizebox{0.98\linewidth}{!}{
\begin{tabular}{ccccccc}
\toprule
 Dataset & {\ours} & {w/o $\ell_{\text{CE}}$} & {w/o $\ell_{\text{CL}}$} & {w/o $\omega_{\bA}$} & {w/o E2E} & {w/o NE}  \\
\midrule
PNC & \textbf{83.4 ± 2.0} & 82.6 ± 1.4 & 81.8 ± 1.5  & 82.6 ± 1.8 & 81.0 ± 1.8 & 71.2 ± 2.9 \\
ABCD & \textbf{94.2 ± 0.2} & 93.7 ± 0.5 & 93.8 ± 0.4  & 93.5 ± 0.3 & 92.8 ± 0.6  & 88.5 ± 2.6 \\ 
\bottomrule
\end{tabular}
}
\caption{Ablation study: AUROC performance on two datasets after removing  specific components. }
\label{tab:abla}
\end{table}


\begin{figure}[!t]
	\centering
	\vspace{-1ex}
	\subfloat[PNC]{
		\includegraphics[width=0.46\linewidth]{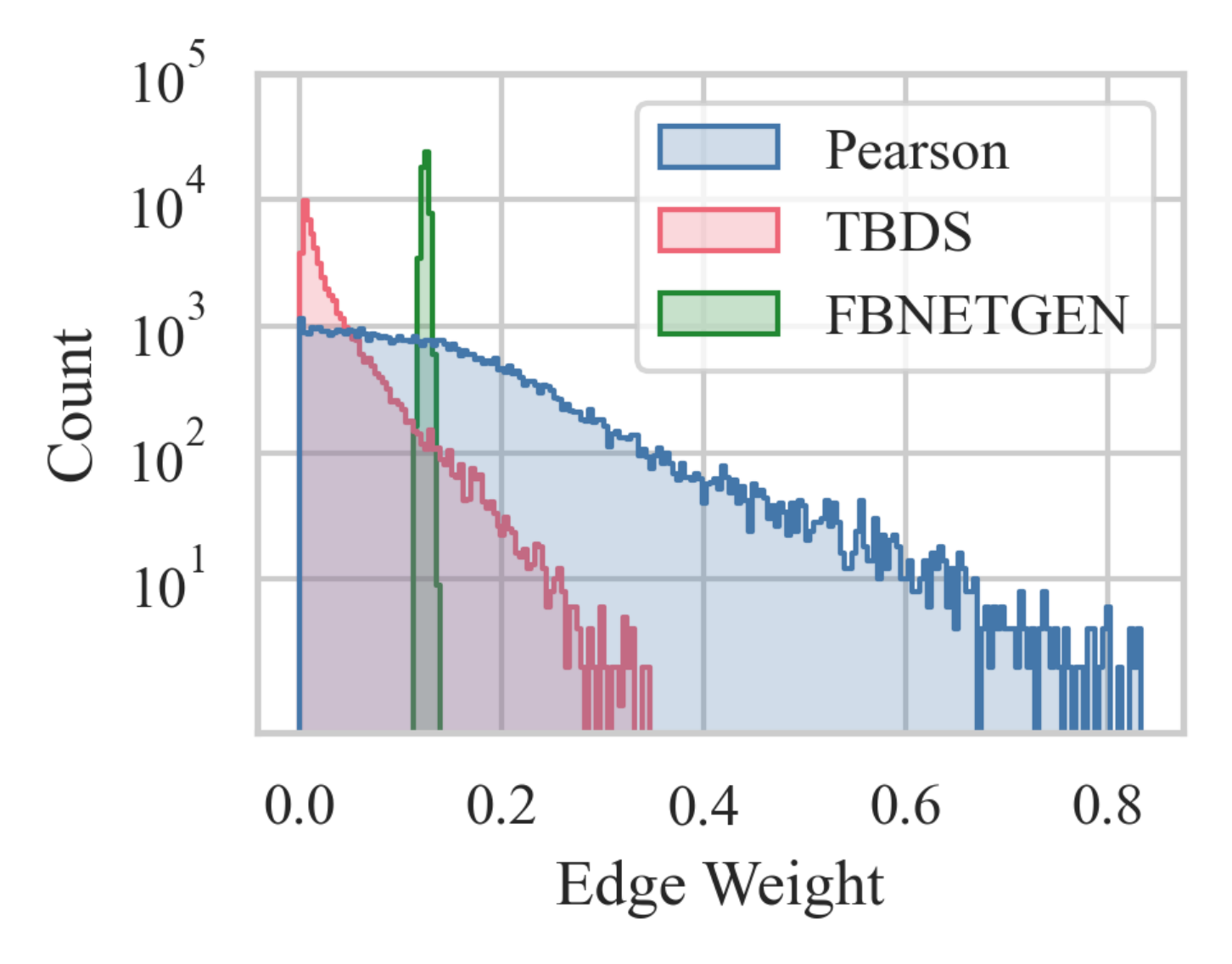}
		\label{fig:distr_pnc}
	} 
	\subfloat[ABCD]{
		\includegraphics[width=0.45\linewidth]{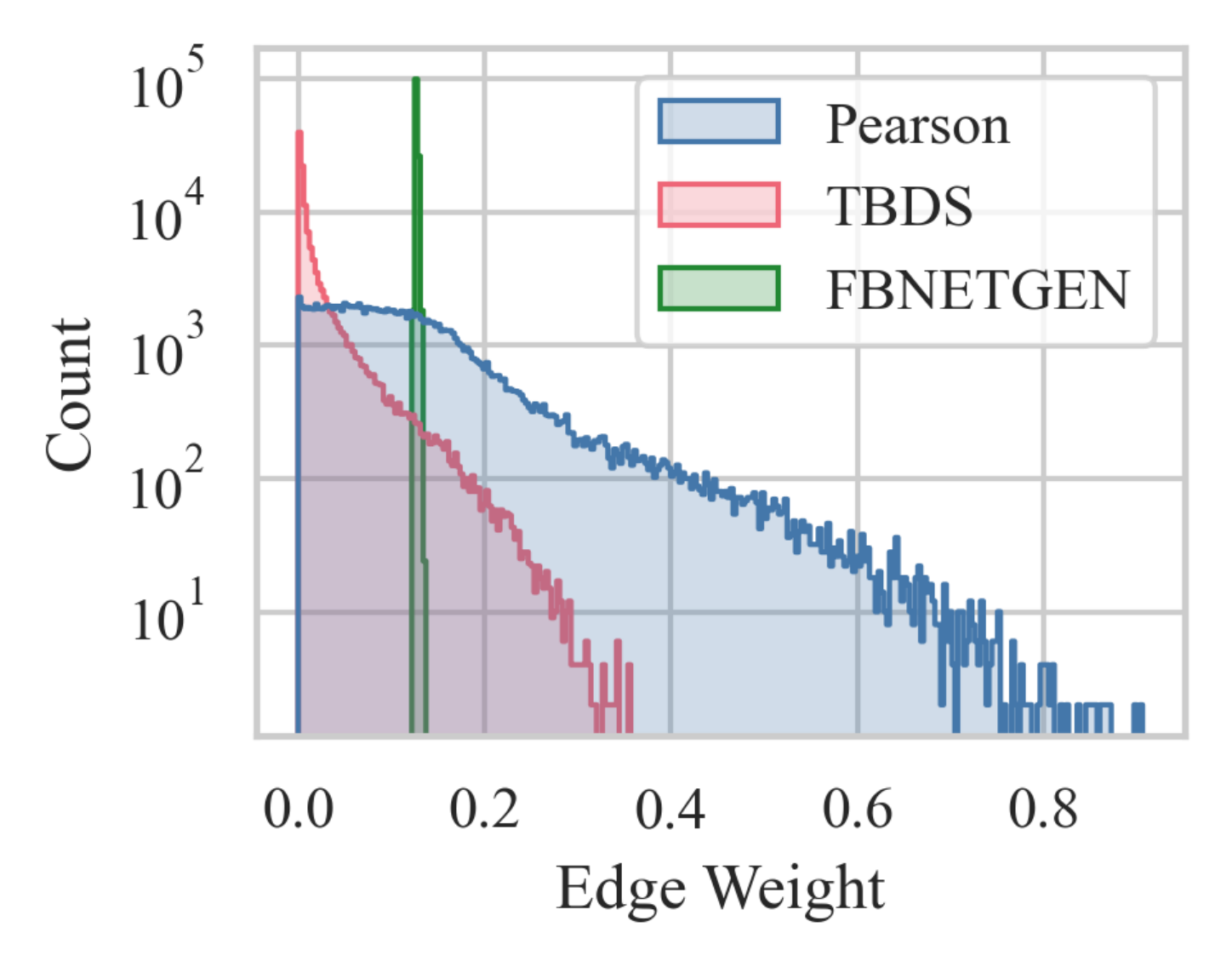}
		\label{fig:distr_abcd}
	}
	\caption{Weight density distributions of the generated brain networks. {\ours} produces much sparser brain connectivities.}
\label{fig:distribution}
\end{figure}

\begin{figure*}[t]
	\centering
	\subfloat[{\ours} on PNC]{
		\includegraphics[width=0.23\linewidth]{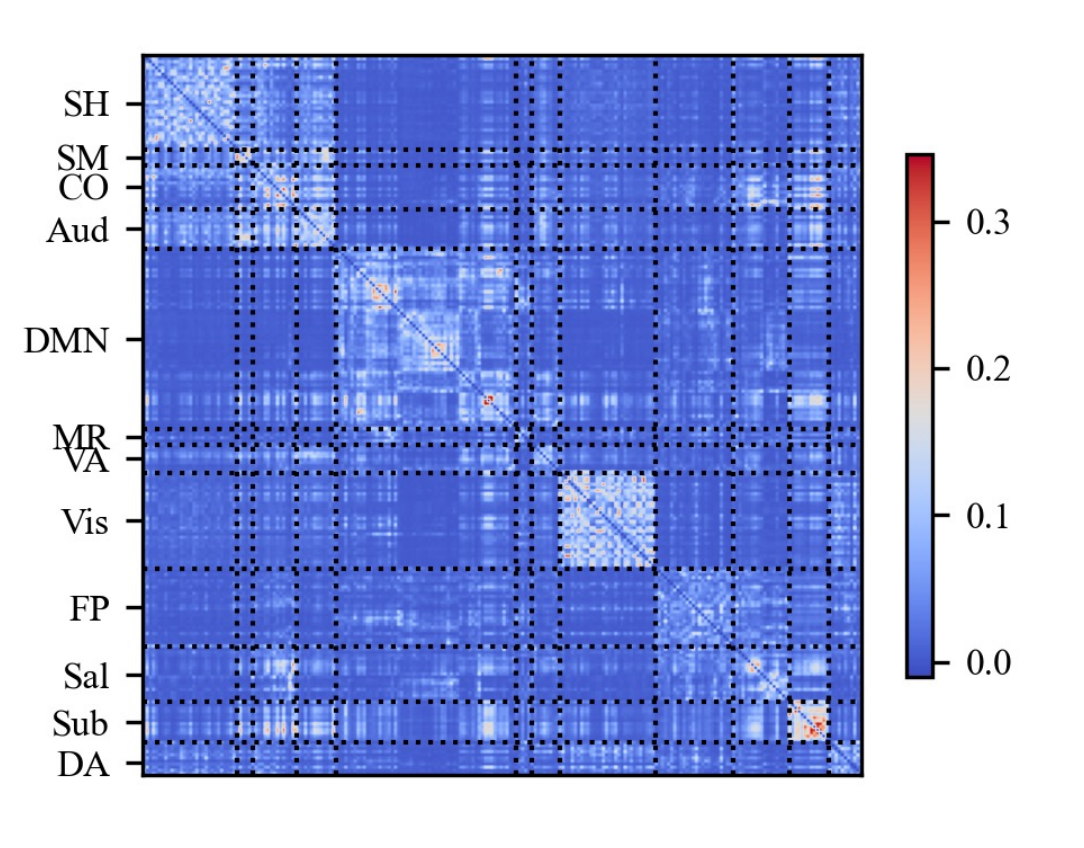}
		\label{fig:tbds_pnc}
	} 
	\subfloat[Pearson on PNC]{
		\includegraphics[width=0.23\linewidth]{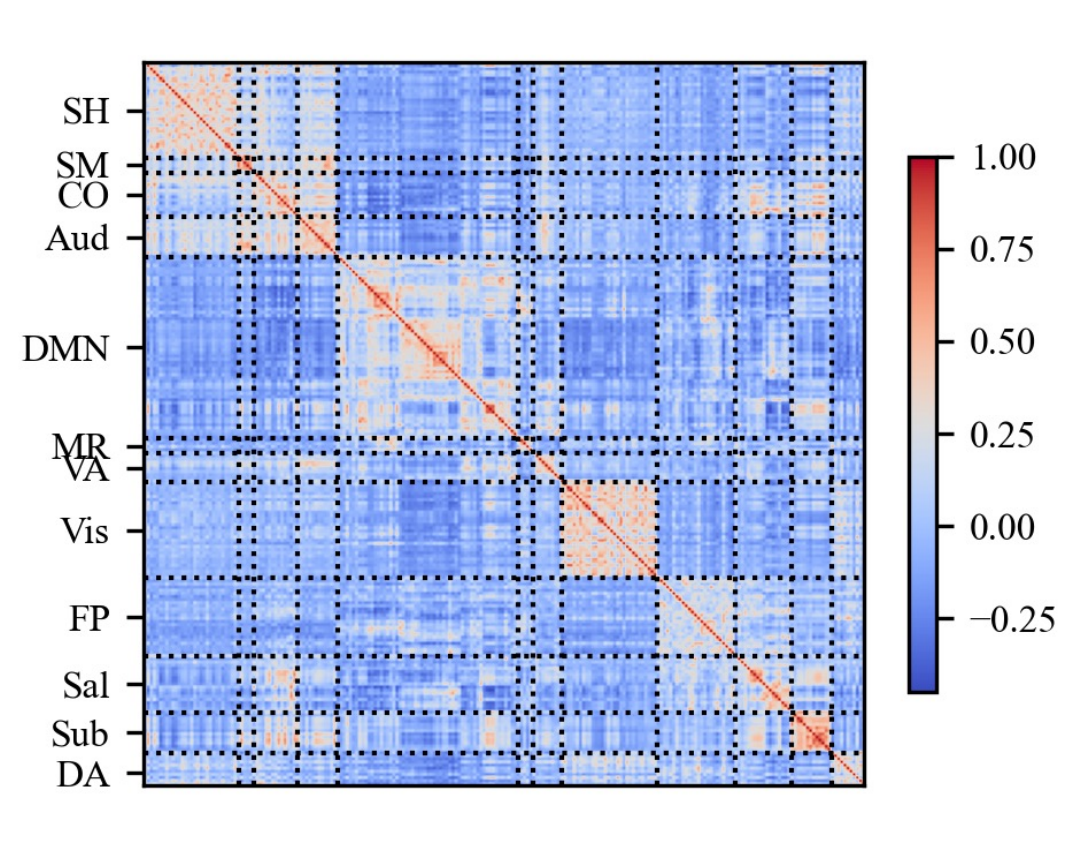}
		\label{fig:pearson_pnc}
	}
	\subfloat[{\ours} on ABCD]{
		\includegraphics[width=0.23\linewidth]{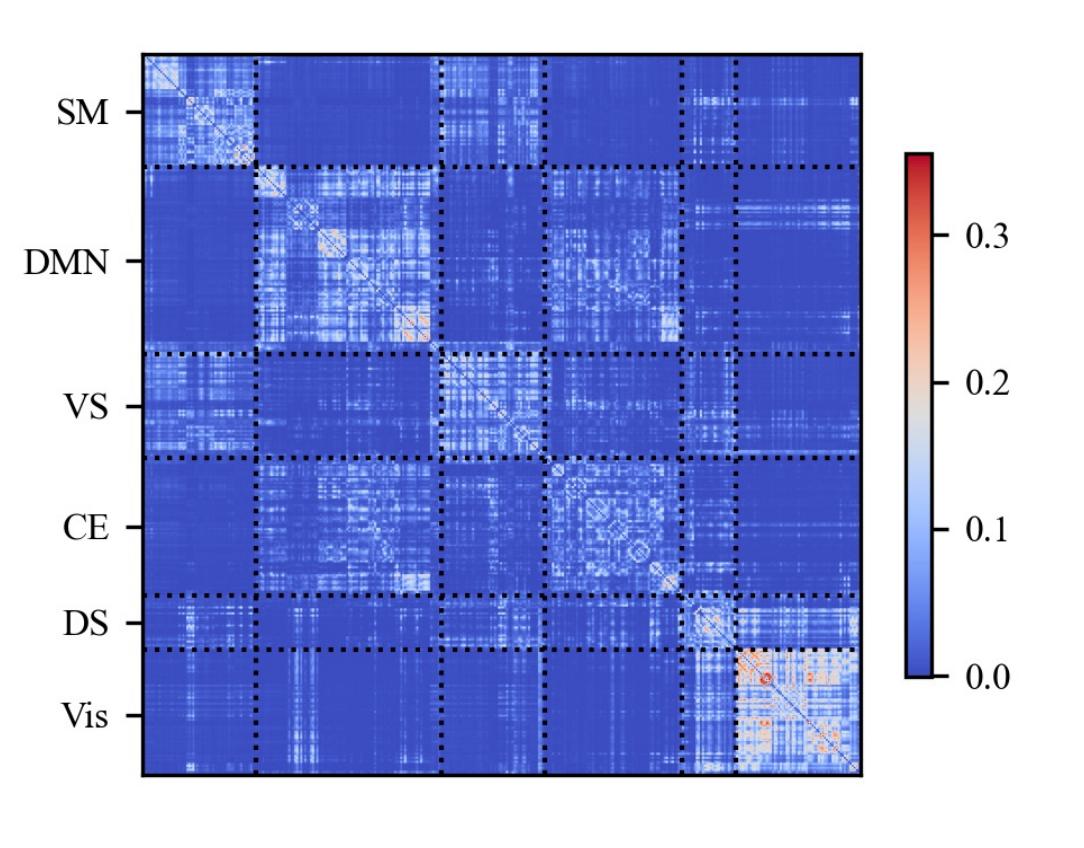}
		\label{fig:tbds_abcd}
	} 
	\subfloat[Pearson on ABCD]{
		\includegraphics[width=0.23\linewidth]{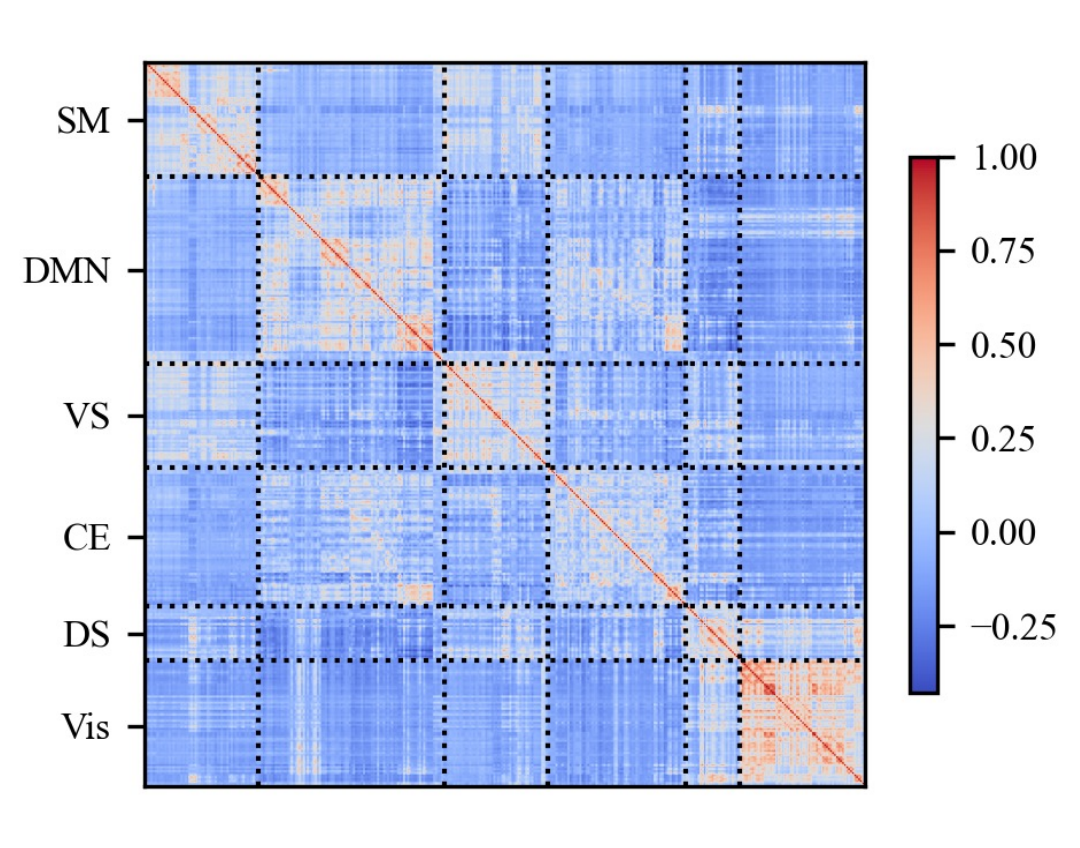}
		\label{fig:pearson_abcd}
	}
	\caption{Visualizations of brain connectivities generated via {\ours} vs. Pearson. Warmer colors indicate higher values. The neural systems on PNC include Somatomotor Hand (SH), Somatomotor Mouth (SM), Subcortical (Sub), Visual (Vis), Auditory (Aud), Cingulo-opercular (CO), Salience (Sal), Default mode (DMN), Fronto-parietal (FP), Ventral attention (VA), Dorsal attention (DA), and Memory retrieval (MR). For ABCD, the neural systems include SM, DMN, Ventral salience (VS), Central executive (CE), Dorsal salience (DS), and Vis.}
\label{fig:connectivity}
\end{figure*}

\begin{figure*}[t]
	\centering
	\subfloat[Male (Label 0)]{
		\includegraphics[width=0.48\linewidth]{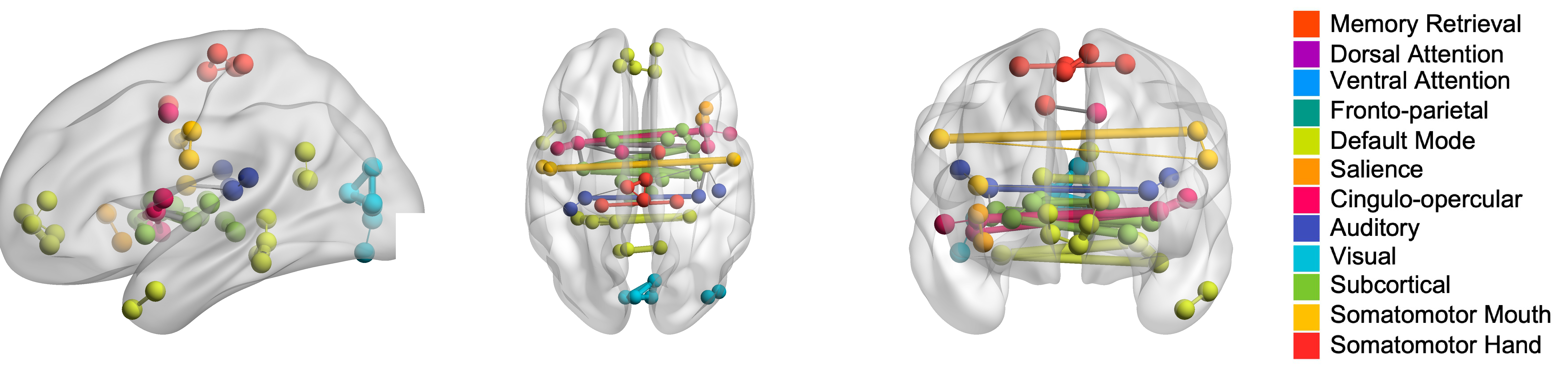}
		\label{fig:label_0}
	} 
	\subfloat[Female (Label 1)]{
		\includegraphics[width=0.48\linewidth]{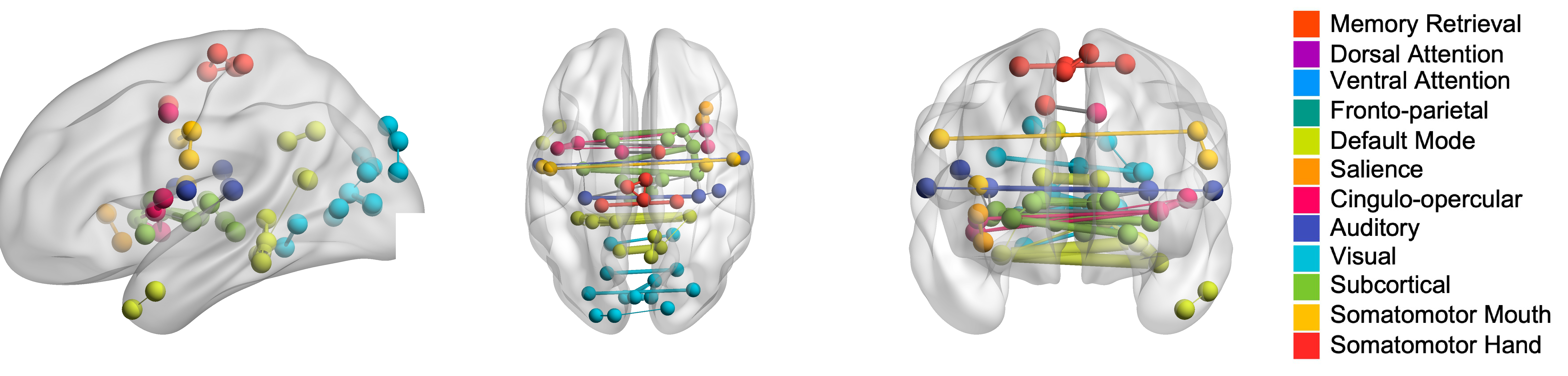}
		\label{fig:label_1}
	}
	\caption{Visualizations of predominant connectivities generated via {\ours} on different biological sexes. Edges spanning multiple neural systems are colored gray, whereas those linking nodes within the same neural system are colored appropriately. Edge widths reflect the connectivity strengths.}
\label{fig:predominant}
\end{figure*}

\subsection{Performance Analysis (RQ2)}
\label{sec:abla}

\noindent  $\diamond$ \textbf{Ablation Studies.} We further examine the designed components in our graph generation (Section \ref{ref:graph_gen}): the cross-entropy loss from the target task $\ell_{\text{CE}}$ (Eq.~\ref{eq:contrast_gen}), the contrastive loss $\ell_{\text{CL}}$  (Eq.~\ref{eq:contrast_gen}), the uncertainty-aware weight for test data $\omega_{\bA}$ (Eq.~\ref{eq:contrast_gen_test}),  
the whole end-to-end learning process (E2E)\footnote{When removing E2E, the model first generates DAG via Eq.~\ref{eq:log_likelihood} to Eq.~\ref{eq:unconstrain} then directly use the DAG for target task prediction via $g_{\phi}$ (Eq.~\ref{eq:gcn}-~\ref{eq:gcn_pred}).}, and the strategy for handling negative weight (Eq.~\ref{eq:gcn}). 
From Table~\ref{tab:abla}, we find that all these components contribute to the final performance. 
Removing $\ell_{\text{CE}}$ and $\ell_{\text{CL}}$ results in performance degradation, as they focus on different parts of the task knowledge (predictions and representations respectively). 
The \emph{uncertainty-based reweighting} $\omega_{\bA}$ prevents the generative model from overfitting to the noise of the target classifier, and leads to overall better performance. 
Combining them in {\ours} provides the best overall effectiveness.
\emph{End-to-end learning} makes our learning framework more flexible, thus leads to 2.2\% performance gain on average.
Our proposed strategy for \emph{tackling negative edges} in Eq.~\ref{eq:gcn} effectively resolves the issue, which  brings significantly performance gain (up to 17\%).


\noindent $\diamond$ \textbf{Sparsity of the Generated DAGs.} Fig.~\ref{fig:distribution} illustrates the connection weight distribution among different ROIs. It is shown that {\ours} fulfills the purpose of sparsity, as most of the weights are close to zero. In contrast, the original Pearson correlation matrix contains many noisy edges and the brain connectivity is dense. 
FBNetGen leverages dot product with softmax normalization to calculate the edge weights, but the softmax operation makes the variance of the edge weights smaller, and the generated graph is not sparse enough.




\subsection{Case Studies (RQ3)}

We explicitly visualize the generated brain connectivities of {\ours} and Pearson in Fig. \ref{fig:connectivity}. It clearly shows that our proposed task-aware DAG learning method generates much sparser connectivities compared with the correlation based method, which effectively controls the time complexity for downstream analysis. At the same time, the edge strength patterns observed in Pearson graphs remain in DAG sparse graphs, indicating that our method can highlight meaningful signals while suppressing random noise that may hinder the identification of potential biomarkers.   

Furthermore, to demonstrate the class discrimination ability of our task-aware DAG brain connectivity, we visualize and compare the strong connections of the learned brain networks between two biological sex groups, as shown in Fig. \ref{fig:predominant}. Specifically, we first divide the learned graphs based on their class labels and calculate an average network by taking the average weight of each edge within the same class. The top 100 most predominant edges are then visualized using the BrainNet Viewer \cite{xia2013brainnet}. By comparing Fig. \ref{fig:predominant}(a) and Fig.  \ref{fig:predominant}(b), we can observe that the main difference in important connections between different biological sex groups lies in the Default Mode Network (DMN), the Auditory Network (Aud) and the Visual Network (Vis), which is in line with previous studies~\cite{genderfunction} that ROIs with significant biological sex differences are located in the DMN and Aud systems. This indicates that our learnable brain connectivity is task-oriented and retains favorable class discrimination ability.

\section{Conclusion}
\label{sec:conclusion}
In this paper, we propose {\ours}, a task-aware effective brain connectivity generation approach to support clinical predictive tasks. 
In particular, we leverage the DAG structure learning techniques to encode the relations among ROIs, and add two additional regularizations to inject task-specific information.
Our end-to-end framework is efficient with GPU acceleration. 
The experiments on two real-world datasets illustrate the superior performance of {\ours} when compared with advanced baselines, and demonstrates valuable neurological interpretations towards downstream tasks.

\bibliographystyle{IEEEtranS}
\bibliography{ref,midl}

\end{document}